\documentclass[fleqn,twoside]{article}
\usepackage[headings]{espcrc2}

\readRCS
$Id: espcrc2.tex,v 1.2 2004/02/24 11:22:11 spepping Exp $
\usepackage{graphicx, balance}
\usepackage[cmex10]{amsmath}
\usepackage{algorithmic}
\usepackage[dvipsnames,usenames]{color}
\usepackage{array}     	% For tabular alignments
\usepackage{subfigure}
\usepackage[normalem]{ulem}  % for text strikethrough/strikeout
\usepackage[ruled, vlined]{algorithm2e}
\usepackage{verbatim}  	% For multi-line comments
\usepackage{tabularx}  	% for better, more flexible-column tables
\usepackage{multirow}  	% for multirow, multicolumn
\usepackage{booktabs} 	% For very nice tables
\usepackage[table]{xcolor}  % for coloring table rows and columns.
\usepackage[bookmarks]{hyperref}
\usepackage{url}
\usepackage[figuresright]{rotating}
\usepackage{ctable} % for \specialrule command

\title{\textbf{Link-Reliability Based Two-Hop Routing for Wireless Sensor Networks}}

\author{T Shiva Prakash\address[DCSE]{Department of Computer Science and Engineering, University Visvesvaraya College of\\ ~Engineering, Bangalore University, Bangalore 560 001 India, Contact: spt@ieee.org\\},
K B Raja\address[DECE]{Department of Electronics and Communication Engineering, University Visvesvaraya College of \\~Engineering, Bangalore University, Bangalore 560 001 India.\\},
K R Venugopal\address[DCSE]{Department of Computer Science and Engineering, University Visvesvaraya College of \\ ~Engineering, Bangalore University, Bangalore 560 001 India.\\},
S S Iyengar\address[DCSE]{Department of Computer Science, Louisiana State University, Baton Rouge, LA 70803, USA.\\},
L M Patnaik\address{Honorary Professor, Indian Institute of Science, Bangalore 560 012, India}}
\runtitle{Link-Reliability Based Two-Hop Routing for Wireless Sensor Networks}
\runauthor{Shiva Prakash T, et al.,}

\begin{document}
\begin{abstract}

Wireless Sensor Networks (WSNs) emerge as underlying infrastructures for new classes of large scale networked embedded systems. 
However, WSNs system designers must fulfill the Quality-of-Service (QoS) requirements imposed by the applications (and users). 
Very harsh and dynamic physical environments and extremely limited energy/computing/memory/communication node resources 
are major obstacles for satisfying QoS metrics such as reliability, timeliness and system lifetime. The limited communication 
range of WSN nodes, link asymmetry and the characteristics of the physical environment lead to a major source of QoS degradation 
in WSNs. This paper proposes a Link Reliability based Two-Hop Routing protocol for wireless Sensor Networks (WSNs). The protocol achieves to reduce packet deadline miss ratio while considering link reliability, two-hop velocity and power efficiency and utilizes memory and computational effective methods for estimating the link metrics. Numerical results provide insights that the protocol has a lower packet deadline miss ratio and longer sensor network lifetime. The results show that the proposed protocol is a feasible solution to the QoS routing problem in wireless sensor networks that support real-time applications. 
\end{abstract}

% typeset front matter (including abstract)
\maketitle

\section{INTRODUCTION}
\label{section:Introduction}
Wireless Sensor Networks (WSNs) form a framework to accumulate and analyze real time data in smart environment applications. WSNs are composed of inexpensive low-powered micro sensing devices called $motes$ [1], having limited computational capability, memory size, radio transmission range and energy supply. Sensors are spread in an environment without any predetermined infrastructure and cooperate to accomplish common monitoring tasks which usually involves sensing environmental data. With WSNs, it is possible to assimilate a variety of physical and environmental information in near real time from inaccessible and hostile locations. 

\begin{table*}
\centering
\caption{Our Results and Comparison with Previous Results for QoS Routing in Wireless Sensor Networks.}
\begin{center}
\footnotesize	
%\otoprule  %\toprule[2pt]
\begin{tabular}{lllllll}
%\hline
%\specialrule{2.5pt}{1pt}{1pt}
\specialrule{.1em}{.05em}{.05em} 
%\otoprule
\bfseries {Related} & \bfseries {Protocol} & \bfseries {Considered} & \bfseries {Estimation} & \bfseries {Performance} \\ 
\bfseries {Work}    &  \bfseries {Name}    & \bfseries {Metric}	    & \bfseries {Method}     & \bfseries \\
%\otoprule %\toprule[2pt]%\hline
%\hline
\specialrule{.1em}{.05em}{.05em} 
    Tian He	&SPEED (Stateless 	&one-hop delay &EWMA (Exponential &Improves end-to-end delay \\
    \emph{et al.,}[2] &Protocol for  &and residual energy &Weighted Moving &and provides good response\\
				&End-to-End Delay) & &Average) &to congestion.\\
\hline
    E. Felemban & MMSPEED  &one-hop delay, &EWMA (Exponential  &Provides service \\
    \emph{et al.,}[3] &(Multi-path and &link reliability &Weighted Moving &differentiation and\\
    &Multi-SPEED &and residual energy &Average) &QoS guarantee in the\\
    &Routing Protocol) & & &timeliness and reliability\\
    & & & &domains.\\
\hline
    Chipera & RPAR (Real-Time  &one-hop delay and &Jacobson Algorithm &Provides real-time routing\\
    \emph{et al.,}[4] &Power Aware &transmission power & &and dynamic power adaption\\
    &Routing) & & &to achieve low delays\\
    & & & &at optimal energy cost.\\
\hline
    Y. Li & THVR (Two-Hop &two-hop delay and &WMEWMA (Window &Routing decision is made\\
    \emph{et al.,}[5] &Velocity Based &residual energy &Mean Exponential &based on two-hop velocity \\
    &Routing Protocol) & &Weighted Moving &integrated with energy \\
    & & &Average &balancing mechanism which \\ 
    & & & &achieves lower DMR.\\
\hline
    This paper &RTLRR &two-hop delay,&EWMA (Exponential  &The protocol considers link\\  
    &(Real-time  &link reliability &Weighted Moving &reliability and uses\\
    &Link Reliability &and residual energy &Average) and &dynamic velocity as per\\
    &Routing) & &WMEWMA &the desired deadline,\\
    & & & &energy is efficiently\\
    & & & &balanced among the nodes.\\
%\hline
\specialrule{.1em}{.05em}{.05em} 
\end{tabular}
\end{center}
\end{table*}
\vskip 2mm
WSNs have a wide variety of applications in military, industry, environment monitoring and health care. WSNs operate unattended
in harsh environments, such as border protection and battlefield reconnaissance hence help to minimize the risk to human life.  used of WSNs are used extensively in the industry for factory automation, process control, real-time monitoring of machines, detection of radiation and leakages and remote monitoring of contaminated areas, aid in detecting possible system deterioration and to initiate precautionary maintenance routine before total system breakdown. WSNs are being rapidly deployed in patient health monitoring in a hospital environment, where different health parameters are obtained and forwarded to health care servers accessible by medical staff and surgical implants of sensors can also help monitor a patient’s health. 
\vskip 2mm
Emerging WSNs have a set of stringent QoS requirements that include timeliness, high reliability, availability and integrity. The competence of a WSN lies in its ability to provide these QoS requirements. The timeliness and reliability level for data exchanged between sensors and control station is of paramount importance especially in real time scenarios. The deadline miss ratio (DMR) [6], defined as the ratio of packets that cannot meet the deadlines should be minimized. Sensor nodes typically use batteries for energy supply. Hence, energy efficiency and load balancing form important objectives while designing protocols for WSNs. Therefore, providing corresponding QoS in such scenarios pose to be a great challenge. Our proposed protocol is motivated primarily by the deficiencies of the previous works (explained in the Section 2) and aims to provide better Quality of Service.
\vskip 2mm
This paper explores the idea of incorporating QoS parameters in making routing decisions $i.e.$,: (i) reliability  (ii) latency and (iii) energy efficiency. Traffic should be delivered with reliability and within a deadline. Furthermore, energy efficiency is intertwined with the protocol to achieve a longer network lifetime. Hence, the protocol is named, Link Reliability based Two-Hop Routing (LRTHR). The protocol proposes the following features.

\begin{enumerate}
\item
Link reliability is considered while choosing the next router, this selects paths which have higher probability of successful delivery.
\item
Routing decision is based on two-hop neighborhood information and dynamic velocity that can be modified according to the required deadline, this results in significant reduction in end-to-end DMR (deadline miss ratio).
\item
Choosing nodes with higher residual energy balances, the load among nodes and results in prolonged lifetime of the network.
\end{enumerate}
The proposed protocol is devised using a modular design, separate modules are dedicated to each QoS requirement. The
link reliability estimation and link delay estimation modules use memory and computational effective methods
suitable for WSNs. The node forwarding module is able to make the optimal routing decision using the estimated metrics. We test the performance of our proposed approaches by implementing our algorithms using $ns$-$2$ simulator. Our results
demonstrates the performance and benefits of LRTHR over earlier algorithms.
\vskip 2mm
The rest of the paper is organized as follows: Section 2 gives a review of Related Works. Section 3 and Section 4 explains the Network Model, notations, assumptions and working of the algorithm. Section 5 is devoted to the Simulation and Evaluation of the algorithm. Conclusions are presented in Section 6.
% Section 5 presents the mathematical analysis of the solution.

%%%%%%%%%%%%%%%%%%%%%%%%%%%%%%%%%%%%%%%%%%%%%%%%%%%%%%%%%%%%%%%%%%%%%%%%%%%%%%%%%%%%%%%%%%%%%%%%%%%%%%%%%%%%%%%%%%%%%%%%%%%

\section{RELATED WORK}
\label{section:related work}
Stateless routing protocols which do not maintain per-route state is a favorable approach for WSNs. The idea of stateless routing is to use location information available to a node locally for routing, i.e., the location of its own and that of its one-hop neighbors without the knowledge about the entire network. These protocols scale well in terms of routing overhead because the tracked routing information does not grow with the network size
or the number of active sinks. Parameters like distance to sink, energy efficiency and data aggregation, need to be considered to select the next router
among the one-hop neighbors.
\vskip 2mm
SPEED (Stateless Protocol for End-to-End Delay) [2] is a well known stateless routing protocol for real-time communication in sensor networks. It is based on geometric routing protocols
such as greedy forwarding GPSR (Greedy Perimeter State Routing) [7][8]. It uses non-deterministic forwarding to balance each flow among multiple 
concurrent routes. SPEED combines Medium Access Control (MAC) and network layer mechanism to maintain a uniform speed across the 
network, such that the delay a packet experiences is directly proportional to its distance to the sink. At the MAC layer, a single hop relay 
speed is maintained by controlling the drop/relay action in a neighbor feedback loop. Geographic forwarding is used to route data to its destination selecting the next hop as a neighbor from the set of those with a relay speed higher that the desired speed. A back pressure re-routing mechanism is employed to re-route traffic around congested areas if necessary.
\vskip 2mm
 Lu \emph{et al.}, [9] describe a packet scheduling policy, called Velocity Monotonic Scheduling, which inherently accounts for both time and distance constraints.
Sequential Assignment Routing (SAR) [10] is the first routing protocol for sensor networks that creates multiple trees routed from one-hop neighbors of the sink by taking into consideration both energy resources, QoS metric on each path and priority level of each packet. However, the protocol suffers from the overhead of maintaining the tables and states at each sensor node especially when the number of nodes is large.
\vskip 2mm
MMSPEED (Multi-path and Multi-SPEED Routing Protocol) [3] is an extension of SPEED that focuses on differentiated QoS options for real-time applications with multiple different deadlines. It provides differentiated QoS options both in timeliness domain and the reliability domain. For timeliness, multiple QoS levels are supported by
providing multiple data delivery speed options. For reliability, multiple reliability requirements are supported by probabilistic multi-path forwarding. The protocol provides end-to-end QoS provisioning by employing localized geographic forwarding using immediate neighbor information without end-to-end path discovery and maintenance. It utilizes dynamic compensation which compensates for inaccuracy of local decision as a packet travels towards its destination. The protocol adapts to network dynamics. MMSPEED does not include energy metric during QoS route selection.  Chipera \emph{et al.},[4](RPAR:Real-Time Power Aware Routing) have proposed another variant of SPEED. Where a node will change its transmission power by the progress towards destination and packet's slack time in order to meet the required velocity; they have not considered residual energy and reliability. 
\vskip 2mm
Mahapatra \emph{et al.},[11] assign an urgency factor to every packet depending on the residual distance and time the packet neesds to travel, and determines the distance the packet needs to be forwarded closer to the destination to meet its deadline. Multi-path routing is performed only at the source node for increasing reliability. Some routing protocols with congestion awareness have been proposed in [12][13]. Other geographic routing protocols such as [14-17] deal only with energy efficiency
and transmission power in determining the next router. Seada \emph{et al.},[18] proposed the PRR (Packet Reception Rate) $\times$ Distance greedy forwarding 
that selects the next forwarding node by multiplying the PRR by the distance to the destination. Recent geographical routing protocols have been proposed, such as DARA (Distributed Aggregate Routing Algorithm) [19], GREES (Geographic Routing with Environmental Energy Supply) [20], DHGR (Dynamic Hybrid Geographical Routing) [21] and EAGFS (Energy Aware Geographical Forwarding Scheme) [22]. They define either the same combined metric (of all the considered QoS metrics) [2][22][20] or several services but with respect to only one metric [14][13].
\vskip 2mm
Sharif \emph{et al.},[23] presented a new transport layer protocol that prioritizes sensed information based on its nature while simultaneously supporting the data reliability and congestion control features. Rusli \emph{et al.},[24] propose an analytical framework model based on Markov Chain of OR and M/D/l/K queue to measure its performance in term of end-to-end delay and reliability in WSNs. 
Koulali \emph{et al.},[25] propose a hybrid QoS routing protocol for WSNs based on a customized Distributed Genetic Algorithm (DGA) that accounts for delay and energy constraints. Yunbo Wang \emph{et al.},[26] investigate the end-to-end delay distribution, they develop a comprehensive cross-layer analysis framework, which employs a stochastic queueing model in realistic channel environments. Ehsan \emph{et al.},[27] propose energy and cross-layer aware routing schemes for multichannel access WSNs that account for radio, MAC contention and network constraints.
\vskip 2mm
All the above routing protocols are based on one-hop neighborhood information. However, it is expected that multi-hop information can lead to improved performance in many issues including message broadcasting and routing. Spohn \emph{et al.},[28] propose a localized algorithm for computing two-hop connected dominating set to reduce the number of redundant broadcast transmissions. An analysis in [29] shows that in a network of $n$ nodes of total of $O(n)$ messages are required to obtain 2-hop neighborhood information and each message has $O(log n)$ bits. 
Chen \emph{et al.},[30] study the performance of 1-hop, 2-hop and 3-hop neighborhood information based
routing and propose that gain from 2-hop to 3-hop is relatively minimal, while that from 1-hop to 2-hop based routing is significant.
\vskip 2mm
Li \emph{et al.},[5] have proposed a Two-Hop Velocity Based Routing Protocol (THVR). The routing choice is decided on the two-hop relay velocity and residual energy, an energy efficient packet drop control is included to enhance packet utilization efficiency while keeping low packet deadline miss ratio. However, THVR does not consider reliability while deciding the route. The protocol proposed in this paper is different from THVR. It considers reliability and uses dynamic velocity that can be altered for each packet as per the desired deadline. It considers energy efficiently and balances the load only among nodes estimated to offer the required QoS.

\section{PROBLEM DEFINITION}
\label{section:pd}
The topology of a wireless sensor network may be described by a graph $G=(N,L)$, where $N$ is the set of nodes and $L$ is the set of links. The objectives are to,
\begin{itemize}
\item Minimize the deadline miss ratio (DMR).
\item Reduce the end-to-end packet delay.
\item Improve the energy efficiency (ECPP-Energy Consumed Per Packet) of the network.
\end{itemize}
\subsection{Network Model and Assumptions}
In our network model, we assume the following:
\begin{itemize}
  \item The wireless sensor nodes consists of $N$ sensor nodes and a sink, the sensors are distributed randomly in a field.
  \item The nodes are aware of their positions through internal global positioning system (GPS), so each sensor has a estimate of its current position.
  \item The $N$ sensor nodes are powered by a non renewable on board energy source. When this energy supply is exhausted the sensor becomes non-operational. All 
  nodes are supposed to be aware of their residual energy and have the same transmission power range.
  \item The sensors share the same wireless medium each packet is transmitted as a local broadcast in the neighborhood. The sensors are neighbors if they are in
  the transmission range of each other and can directly communicate with each other. We assume a MAC protocol, $i.e.$, IEEE 802.11 which ensures that among
  the neighbors in the local broadcast range, only the intended receiver keeps the packet and the other neighbors discard the packet.
  \item Like all localization techniques, [2][3][31][32][33] each node needs to be aware of its 
  neighboring nodes current state (ID, position, link reliability, residual energy etc), this is done via HELLO messages. 
  \item Nodes are assumed to be stationary or having low mobility, else additional HELLO messages will be needed
  to keep the nodes up-to-date about the neighbor nodes.
  \item In addition, each node sends a second set of HELLO messages to all its neighbors informing them about its one-hop neighbors. 
  Hence, each node is aware of its one-hop and two-hop neighbors and their current state.
  \item The network density is assumed to be high enough to prevent the void situation.
\end{itemize}

%%%%%%%%%%%%%%%%%%%%%%%%%%%%%%%%   WORKING TABLE   %%%%%%%%%%%%%%%%%%%%%%%%%%%%%%%%%%%%%%%%%
%\renewcommand{\multirowsetup}{\centering}
%\newcommand {\otoprule}{\midrule [\heavyrulewidth]}  % This is for getting a bold horizontal line in the middle of the table.
\newcolumntype{A}{>{$}c<{$}}
\newcolumntype{B}{>{$}l<{$}}
\newcolumntype{C}{m{5cm}}
\begin{table}[t]
 		\caption{Notations used in Section 4}
\footnotesize																										 
 	\begin{tabular}{BC}
    \toprule[0.5pt]   %\otoprule
    \multicolumn{1}{b{1.2cm}}{\bfseries Symbol}
    & \multicolumn{1}{b{3cm}}{\bfseries Definition}\\ % width changes the column header field width.
    \toprule[0.5pt]   %\otoprule

    \raggedright N
    & Set of Nodes in the WSN\\

    \raggedright D
    & Destination Node \\
    
    \raggedright S
    & Source Node \\
    
    \raggedright dist(x,y)
    & Distance between a node pair $x,y$ \\
    
    \raggedright N_{1}(x)
    & Set of one-hop Neighbors of node $x$ \\

    \raggedright N_{2}(x)
    & Set of two-hop Neighbors of node $x$ \\
    
    \raggedright F_{1}^{+p}(x)
    & Set of node $x$'s one-hop favorable \\
    
    \raggedright
    & forwarders providing positive progress \\

    \raggedright
    & towards the destination D \\

    \raggedright F_{2}^{+p}(x,y)
    & Set of node $x$'s two-hop \\

    \raggedright
    & favorable forwarders \\

    \raggedright delay_{xy}
    & Estimated hop delay between $x$ and $y$ \\

    \raggedright t_{req}
    & Time deadline to reach Destination D\\
    
    \raggedright V_{req}
    & Required end-to-end packet delivery \\
    
    \raggedright
    & Velocity for deadline $t_{req}$ \\

    \raggedright V_{xy}
    & Velocity offered by $y$ $\in$ $F_{1}^{+p}(x)$  \\

    \raggedright V_{xy \rightarrow z}
    & Velocity offered by $y$ $\in$ $F_{2}^{+p}(x,y)$  \\

    \raggedright S_{req}
    & Node pairs satisfying $V_{xy\rightarrow z} \geq V_{req}$   \\
    
    \raggedright E_{y}^{0}
    & Initial energy of node $y$ \\
    
    \raggedright E_{y}
    & Remaining energy of node $y$ \\
    
    \raggedright prr_{xy}
    & Packet Reception Ratio of link relaying \\

    \raggedright
    & node $x$ to node $y$ \\
    
    \raggedright \alpha
    & Tunable weighting coefficient for delay \\

    \raggedright
    & estimation \\
    
    \raggedright \beta
    & Tunable weighting coefficient for $prr$ \\
   
    \raggedright
    & estimation\\
    
    \raggedright A
    & PRR weight factor \\
    
    \raggedright B
    & Velocity weight factor \\    

    \raggedright C
    & Energy weight factor \\
    
   \raggedright rve_{y \rightarrow z}
    & Reliability, Velocity \\
    
    \raggedright
    & and Energy shared metric\\
    
    \toprule[0.5pt] %\bottomrule
  \label{table:notations}
  \end{tabular}
\end{table}
%%%%%%%%%%%%%%%%%%%%%%%%%%%%%%%%%%%%%%%%%%%%%%%%%%%%%%%%%%%%%%%%%%%%%%%%%%%%%%%%%%%%%%%%%%%%%%%%%

\section{ALGORITHM}
\label{section:algo}

LRTHR has three components: a link reliability estimator, a delay estimator, a node forwarding metric incorporated with the 
dynamic velocity assignment policy. The proposed protocol LRTHR implements the modules for estimating transmission delay and packet delivery ratios using efficient methods. The packet delay is estimated at the node itself and the packet delivery ratio is estimated by the neighboring nodes. These parameters are updated on reception of a HELLO packet, the HELLO messages are periodically broadcast to update the estimation parameters. The overhead caused by the 1-hop and 2-hop updating are reduced by piggybacking the information 
in ACK, hence improving the energy efficiency. The notations used in this paper are given in Table \ref{table:notations}. The protocol is based on the following parameters: (i) Link Reliability Estimation; (ii) Link Delay Estimation; and (iii) Node Forwarding Metric

\subsection{Link Reliability Estimation}
The Packet Reception Ratio (PRR) of the link relaying node $x$ to $y$ is denoted by $prr_{xy}$. It denotes the probability of successful 
delivery over the link. Window Mean Exponential Weighted Moving Average (WMEWMA) based link quality estimation is used for the proposed protocol. The window mean exponential weighted moving average estimation applies filtering on PRR, thus providing a metric that resists transient fluctuations of PRR, yet is responsive to major link quality changes.
\vskip 2mm
This parameter is updated by node $y$ at each window and inserted into the HELLO message packet for usage by node $x$ in the next window. 
Eq. \ref{eqn:Eq1} shows the window mean exponential weighted moving average estimation of the link reliability, $r$ is the number of packets received, $m$ is the number of packets missed and $\alpha \in [0,1]$ is the history control factor, which controls the effect of the previously estimated value on the new one, $\frac{r}{r+m}$ is the newly measured PRR value.

%%%%%%%%%%%%%%%% LR2HR Algorithm %%%%%%%%%%%%%%%%%%%
\begin{algorithm}
\begin{footnotesize}
\KwIn{$x$, $D$, $F_1^{+p}(x)$, $F_2^{+p}(x)$, $lt$}
\KwOut{Node $y$ providing positive progress towards D}
\BlankLine
%$V_{req}$ = $\frac{dist(x,D)}{lt}$ \;
%\For{each $y$ $\in$ $F_2^{+p}(x)$}{
%	$V_{xy \rightarrow z}$ = $\frac{dist(x,D) - dist(k,D)}{$delay_{xy}$ + $delay_{yz}$}$ \;
%}

$V_{req}$ = $\frac{dist(x,D)}{lt}$ \;
\For{each $y$ $\in$ $F_2^{+p}(x)$}{
%\For{$i = 0$ \emph{\KwTo} $n_{gk}$}{
	$V_{xy \rightarrow z}$ = $\frac{dist(x,D) - dist(k,D)}{delay_{xy} + delay_{yz}}$ \;
}
$S_{req}$ = \{$F_2^{+p}(x)$ : $V_{xy\rightarrow z} \geq V_{req}$\} \;
\If{($\mid$ $S_{req}$ $\mid$) = 1}{
	return $y$ $\in$ $S_{req}$\;
}
\Else{	
	\For{each $y$ $\in$ $S_{req}$}{
 		$rve_{xy\rightarrow z}$ =
		A $\times$ $\frac{prr_{xy}}{\displaystyle \sum_{y \in S_{req}}(prr_{xy})}$ +		
		B $\times$ $\frac{V_{xy\rightarrow z}}{\displaystyle \sum_{y \in S_{req}}V_{xy\rightarrow z}}$ + 		
		C $\times$ $\frac{E_y/E_y^0}{\displaystyle \sum_{y \in S_{req}}(E_y/E_y^0)}$ \;
		Find $y$ with $Max$ \{$rve_{xy\rightarrow z}\}$ \;
	}
}
	\Return $y$ $\in$ $S_{req}$\;
\caption{Link Reliability Based Two-Hop Routing (LRTHR)}
\label{algo:lr2halgo}
\end{footnotesize}
\end{algorithm}
%%%%%%%%%%%%%%%%%%%%%%%%%%%%%%%%%%%%%%%%%%%%%%%%%%%%%%%%%

\begin{equation}
prr_{xy} = \alpha \times prr_{xy} + (1-\alpha) \times \frac{r}{r+m} 
\label{eqn:Eq1}
\end{equation}
The PRR estimator is updated at the receiver side for each $w$ (window size) received packets, the computation complexity of
this estimator is $O(1)$. The appropriate values for $\alpha$ and $w$ for a stable window mean exponential weighted moving average are $w=30$ and $\alpha=0.5$[34].

\subsection{Link Delay Estimation}
The delay indicates the time spent to send a packet from node $x$ to its neighbor $y$, it is comprised of the queuing delay (delay$_Q$), 
contention delay (delay$_C$) and the transmission delay (delay$_T$).
\begin{equation}
delay_{xy} = delay_{Q} + delay_{C} + delay_{T}
\label{eqn:Eq2}
\end{equation}
If $t_{s}$ is the time the packet is ready for transmission and becomes head of transmission queue, $t_{ack}$ the time of the reception of acknowledgment, $BW$ the network bandwidth and size of the acknowledgment then, $t_{ack} - sizeof(ACK)/BW - t_{s}$ is the recently
estimated delay and $\beta \in [0,1]$ is the tunable weighting coefficient. Eq. \ref{eqn:Eq3} shows the EWMA (Exponential Weighted 
Moving Average) update for delay 
estimation, which has the advantage of being simple and less resource demanding. 
\begin{equation}
\begin{split}
delay_{xy}& = \beta \times delay_{xy} + (1-\beta) \times(t_{ack} - \\
          & \quad sizeof(ACK)/BW - t_{s})
\label{eqn:Eq3}
\end{split}
\end{equation}
$delay_{xy}$ includes estimation of the time interval from the packet that becomes head of line of $x$'s transmission queue until 
its reception at node $y$. This takes into account all delays due
to contention, channel sensing, channel reservation (RTS/CTS) if any, depending on the medium access control (MAC) protocol, time slots etc. The computation complexity of this estimator is $O(1)$. The delay information is further exchanged among two-hop neighbors.

\subsection{Node Forwarding Metric}
In the wireless sensor network, described by a graph $G=(N,L)$. If node $x$ 
can transmit a message directly to node $y$, the ordered pair is an element of $L$. We define for each node $x$ the set $N_1(x)$, which contains the nodes
in the network $G$ that are one-hop $i.e.$, direct neighbors of $x$.
\begin{equation}
N_1(x) = \{y:(x;y) \in E \hspace{0.1cm} and \hspace{0.1cm} y \neq x\}
\label{eqn:Eq4}
\end{equation}

Likewise, the two-hop neighbors of $x$ is the set $N_2(x)$ $i.e.$,
\begin{equation}
N_2(x) = \{z:(y;z) \in E \hspace{0.1cm} and \hspace{0.1cm} y \in N_1(x), \hspace{0.1cm} z \neq x\}
\label{eqn:Eq5}
\end{equation}
The euclidean distance between a pair of nodes $x$ and $y$ is defined by $dist(x,y)$. We define $F_{1}^{+p}(x)$ as the set of $x$'s one-hop favorable forwarders providing positive progress towards the destination $D$. It consists of nodes that are closer to the destination than $x$, $i.e.$,
\begin{equation}
\begin{split}
F_1^{+p}(x)& = \{y \in N_1(x): \\ 
	   & \quad dist(x,D) \hspace{0.1cm} - \hspace{0.1cm} dist(y,D) > 0\}
\label{eqn:Eq6}
\end{split}
\end{equation}
$F_{2}^{+p}(x)$ is defined as the set of two-hop favorable forwarders $i.e.$,
\begin{equation}
\begin{split}
F_2^{+p}(x)& = \{y\in F_1^{+p}(x),z\in N_1(y):\hspace{0.1cm} \\
           & \quad dist(y,D)\hspace{0.1cm}-\hspace{0.1cm} dist(z,D)>0\}
\label{eqn:Eq7}
\end{split}
\end{equation}
We define two velocities; the required velocity $V_{req}$ and the velocity offered by the two-hop favorable forwarding pairs. In SPEED, 
the velocity provided by each of the forwarding nodes in ($F_1^{+p}(x)$) is.
\begin{equation}
V_{xy} = \frac{dist(x,D) - dist(y,D)}{delay_{xy}}
\label{eqn:Eq8}
\end{equation}
As in THVR, by two-hop knowledge, node $x$ can calculate the velocity offered by each of the two-hop favorable forwarding pairs ($F_1^{+p}(x)$,$F_2^{+p}(x)$) $i.e.$,
\begin{equation}
V_{xy\rightarrow z} = \frac{dist(x,D) - dist(z,D)}{delay_{xy} + delay_{yz}}
\label{eqn:Eq9}
\end{equation}
Where, $y \in F_1^{+p}(x)$ and $z \in F_2^{+p}(x)$. The required velocity is relative to the progress made towards the destination
[4] and the time remaining to the deadline, $lt$ (lag time). The lag time is the time remaining until the packet deadline expires. At each hop, the transmitter renews this
parameter in the packet header $i.e.$,
\begin{equation}
lt = lt_{p} - (t_{tx} - t_{rx} + sizeof(packet)/BW)
\label{eqn:Eq10}
\end{equation}
Where $lt$ is the time remaining to the deadline ($t_{req}$), $lt_{p}$ is the previous value 
of $lt$, ($t_{tx} - t_{rx} + sizeof(packet)/BW$) accounts for the delay 
from reception of the packet until transmission. On reception of the packet the node $x$, uses $lt$ to calculate the 
required velocity $V_{req}$ for all nodes in ($F_1^{+p}(x)$,$F_2^{+p}(x)$) as show in Eq. \ref{eqn:Eq11}.
\begin{equation}
V_{req} = \frac{dist(x,D)}{lt}
\label{eqn:Eq11}
\end{equation}
The node pairs satisfying $V_{xy\rightarrow z} \geq V_{req}$ form the set of nodes $S_{req}$. For the set $S_{req}$ we calculate 
the shared metric ($rve_{xy\rightarrow z}$), incorporating the node's link reliability, velocity towards destination and 
remaining energy level of neighbors in  $S_{req}$, as show in Eq. \ref{eqn:Eq12}.
\begin{equation}
\begin{split}
rve_{xy\rightarrow z} = &A \times \frac{prr_{xy}}{\displaystyle \sum_{y \in S_{req}}(prr_{xy})} + \\
	                &B \times \frac{V_{xy\rightarrow z}}{\displaystyle \sum_{y \in S_{req}}V_{xy\rightarrow z}} + \\
                        &C \times \frac{E_y/E_y^0}{\displaystyle \sum_{y \in S_{req}}(E_y/E_y^0)}
\label{eqn:Eq12}
\end{split}
\end{equation}
$A$, $B$ and $C$ are the weighting factors for combining reliability, velocity and energy into the shared metric ($A+B+C=1$). The node $y$ with the 
largest $rve_{xy\rightarrow z}$ will be chosen as the forwarder and the process continuous till the destination is reached. The
Link Reliability Based Two-Hop Routing is shown in Algorithm 1, the computation complexity of this algorithm is $O(F_2^{+p}(x))$. Our proposed protocol is different from THVR, as it considers
reliability and dynamic velocity that can be adjusted for each packet according to the required deadline. It balances the load only among nodes estimated to offer the required QoS.

\begin{figure}
\includegraphics[scale=1.00]{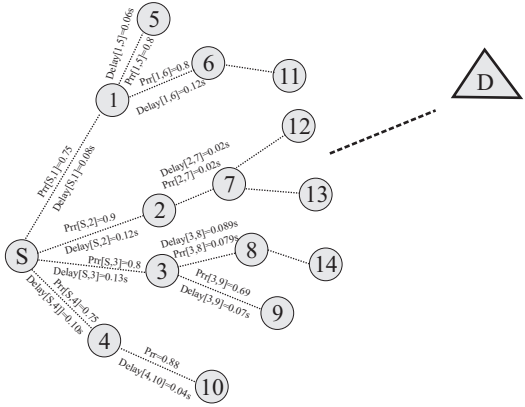}
\begin{center}
\begin{small}
\caption{A Study on Working of LRTHR Protocol}
\end{small}
\end{center}
\end{figure}

\subsection{LRTHR: An Example}
We bring out the working for the proposed protocol in a case study. From Figure 1 if a packet is to be sent from $S$ to $D$, then 
nodes {1,2,3,4} $\in$ $F_{1}^{+p}(S)$. {5,6} $\in$ $F_{1}^{+p}(1)$, 
{7} $\in$ $F_{1}^{+p}(2)$, {10} $\in$ $F_{1}^{+p}(4)$, {14} $\in$ $F_{1}^{+p}(8)$, {12,13} $\in$ $F_{7}^{+p}(7)$, {7} $\in$ $F_{1}^{+p}(2)$. 
The distance between the various nodes and the destination are: (S,$D$)=150m, (1,$D$)=120m, (2,$D$)=108m, (3,$D$)=114m, (4,$D$)=127.5m
(5,$D$)=117m, (6,$D$)=97.5m, (7,$D$)=110m, (8,$D$)=90m, (9,$D$)=97.5m and (10,$D$)=117m. Let the required velocity, 
\begin{equation*}
V_{req} = \frac{150m}{0.55s} = 272.7m/s
\label{eqn:Eq13}
\end{equation*}
Here, the end-to-end deadline is 0.55s. By, Eq. \ref{eqn:Eq8}, each node calculates the velocity ($V_{xy}$) provided by each
of its forwarding nodes in $F_1^{+p}(S)$,
\begin{equation*}
V_{S1} = \frac{150m-120m}{0.08s} = 375m/s
\label{eqn:Eq14}
\end{equation*}
Likewise, the velocity provided by $V_{S2}=(150m-108m)/0.12s=350m/s$, $V_{S3}=(150m-114m)/0.13s=276.92m/s$ 
and $V_{S4}=(150m-127.5m)/0.10s=225m/s$. Thus, from SPEED node 1 has the largest velocity greater than $V_{req}$
and will be chosen as the forwarder and so on.

As per THVR, node $S$ will search among its two-hop neighbors $F_2^{+p}(S)$ $i.e.$, nodes (5,6,7,8,9,10) and calculate
the velocity ($V_{xy\rightarrow z}$) provided by each of the two-hop pairs by Eq. \ref{eqn:Eq9},
\begin{equation*}
\begin{split}
V_{S3\rightarrow 8}& = \frac{dist(S,D) - dist(8,D)}{delay_{S3} + delay_{38}} \\ 
		   & = \frac{150m-90m}{0.13s + 0.079s} \\
		   & = 287.08m/s
\label{eqn:Eq15}
\end{split}
\end{equation*}
Similarly, the velocity provided by the two-hop pairs: $V_{S1\rightarrow 5}$ = $(150m-117m)/(0.11s+0.06s)$ = $235.7m/s$,  
$V_{S1\rightarrow 6}$ = $(150m-97.5m)/(0.08s+0.12s)$ = $262.5m/s$, $V_{S2\rightarrow 7}$ = $(150m-110m)/(0.12s+0.02s)$ = $285.7m/s$,
$V_{S3\rightarrow 9}$ = $(150m-97.5m)/(0.13s+0.07s)$ = $262.5m/s$, $V_{S4\rightarrow 10}$ = $(150m-117m)/(0.10s+0.04s)$ = $235.7m/s$.

The velocity provided by $V_{S3\rightarrow 8}$ is greater than $V_{req}$ and is also the largest among the other two-hop pairs shown
above. Therefore, node 3 will be chosen as the immediate forwarder. But, by LRTHR we also consider the PRR of the links while
choosing the next forwarder, the PRR of link to node 2 is 0.9 and that to link 3 is 0.85, hence node 2 will be chosen 
as the next hop candidate. If the packet arriving at node 2 has taken 0.13ms to travel, then the new deadline to
reach the destination will be 0.42s. The required velocity is updated at node 2 and the next forwarder is 
chosen based on this new value.
\vskip 2mm
In LRTHR, by selecting a link that provides higher PRR, the protocol aids in increasing the probability of successful 
packet delivery to the forwarding node. In THVR, if a path from source to destination has a link with a poor packet
reception ratio, then this may increase the DMR. By, selecting links providing greater PRR on the route, the throughput
(amount of traffic successfully received by the destination) can be increased, obtain a lower DMR, augment the energy
efficiency of the forwarding nodes due to lower number of collisions and re-transmissions. Also, the two-hop neighborhood
information incorporated with the dynamic velocity assignment policy will provide enhanced foresight to the sender in 
identifying the node pair that can provide the largest velocity towards the destination.

\section{PERFORMANCE EVALUATION}

To evaluate the proposed protocol, we carried out a simulation study using $ns$-$2$[35]. The proposed 
protocol (LRTHR) is compared with THVR and SPEED. The simulation configuration consists of 200 nodes 
located in a 200 $m^2$ area. Nodes are distributed following Poisson point process with a node density of 0.005 node/$m^2$. The 
source nodes are located in the region (40m, 40m) while the sink in the area (200m, 200m). The source generated a CBR 
flow of 1 packet/second with a packet size of 150 bytes. 
\vskip 2mm
The MAC layer, link quality and energy consumption parameters are
set as per Mica2 Motes[36] with MPR400 radio as per THVR. Table 3 summarizes the simulation parameters.
THVR and SPEED are QoS protocols and a comparison of DMR (Deadline Miss Ratio), ECPP (Energy Consumed Per Packet $i.e.$, the total
energy expended divided by the number of packets effectively transmitted), the packet average delay (mean of packet delay) and 
worst case delay (largest value sustained by the successful transmitted packet) are obtained.

\begin{table}[ht]
\centering
\caption{Simulation Parameters.}
\begin{tiny}
\begin{center}
\begin{tabular}{|l|c| } \hline
{\bf Simulation Parameters} & {\bf  Value} \\ \hline
{Number of nodes} 		&{200} \\ \hline
{Simulation Topology} 		&{200m x 200m} \\ \hline
{Traffic} 			&{CBR} \\ \hline
{Payload Size} 			&{150 Bytes} \\ \hline
{Transmission Range}			&{40m} \\ \hline
{Initial Battery Energy }		&{2.0 Joules} \\ \hline
{Energy Consumed during Transmit }	&{0.0255 Joule} \\ \hline
{Energy Consumed during Receive }	&{0.021 Joule} \\ \hline
{Energy Consumed during Sleep }		&{0.000005 Joule} \\ \hline
{Energy Consumed during Idle }		&{0.0096 Joule} \\ \hline
{MAC Layer}			&{802.11 with DCF} \\ \hline
{Propagation Model}		&{Free Space} \\ \hline
{Hello Period} 			&{5 seconds} \\ \hline
{PRR - WMEWMA Window} 			&{30} \\ \hline
{PRR - WMEWMA Weight Factor ($\alpha$)}	&{0.6} \\ \hline
{Delay - EWMA Weight Factor ($\beta$)}	&{0.5} \\ \hline
\end{tabular}
\end{center}
\end{tiny}
\end{table}

In the first set of simulations, we consider 10 source nodes with varying deadlines from 100 ms to 700ms. In THVR, the weighting factor $C$ is set at 0.9 to favor end-to-end delay performance, likewise in SPEED we assign $K$=10 for shorter end-to-end delay. In the proposed protocol we set weighting factors ($A$, $B$, $C$) at (0.1, 0.8, 0.1). In each run, 500 packets are transmitted. 

\begin{figure}
\resizebox{7cm}{5cm}{\includegraphics{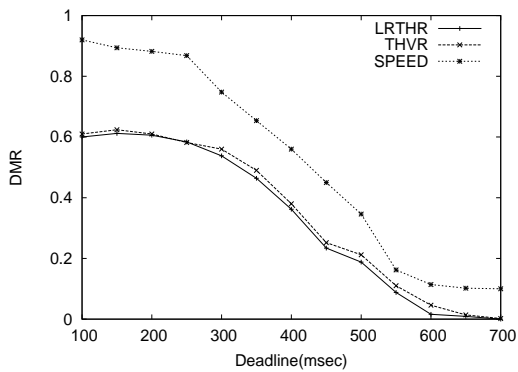}}
\begin{center}
\begin{small}
\caption{DMR vs Deadline}
\end{small}
\end{center}
\end{figure}
\vskip 2mm
Figure 2 illustrates the efficiency of the LRTHR algorithm in reducing the DMR, the DMR characteristics of LRTHR and THVR are similar till a delay of 250ms and the performance of LRTHR is better after that, eventually as the deadlines increase the DMRs converge to zero. 
In comparison, as shown in Figure 2 THVR has a higher DMR, the initiative drop control has a slight negative effect on the DMR. 
\vskip 2mm
In SPEED, when the deadline is stringent (less than 300ms), the SPEED protocol drops packets aggressively at lower deadlines, resulting in a overall higher DMR. Even, when the deadline is 700ms the DMR has not yet converged to zero. The two-hop based routing and dynamic velocity of the LRTHR algorithm is able to aggressively route more packets within the deadline to the sink node, also the protocol is able to select the reliable paths between the sources and the sink, hence it is observed that 
LRTHR has lower DMR than the others in general.

\begin{figure}
\resizebox{7cm}{5cm}{\includegraphics{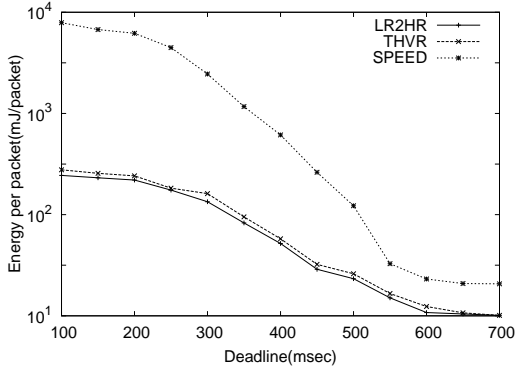}}
\begin{center}
\begin{small}
\caption{ECPP vs Deadline}
\end{small}
\end{center}
\end{figure}
\vskip 2mm

As depicted in Figure 3 the energy consumption per packet successfully transmitted, decreases as the deadline increases. The energy consumption has similar tendency in both LRTHR and THVR but SPEED has a higher energy utilization. The slight variation
of the LRTHR protocol is due to the link reliability incorporated in the route selection which may sometimes select a longer path 
to the destination resulting in higher energy utilization on some paths, but the dynamic velocity will minimize this effect. 
By, selecting links providing higher PRR on the route to the sink, the energy consumption of the forwarding nodes can be 
minimized, due to lower number of collisions and re-transmissions. Also, in the proposed protocol the link delay and 
packet delivery ratios are updated by piggybacking the information in ACK, this will help in reducing the number of feedback
packets and hence reduce the total energy consumed.
\vskip 2mm
In THVR, the initiative drop control module, will drop the packet if it is near the source and cannot meet the required 
velocity, from the perspective of energy utilization. But, in the proposed protocol the packet will not be dropped since the dynamic velocity approach will aid in ensuring that the packet eventually meets the deadline, more packets will be forwarded to the destination and will improve the ECPP. Generally, LRTHR has a lower energy consumption level compared to the other protocols.
\vskip 2mm
Figure 4 compares the packet end-to-end average and worst-case delays respectively. It is observed that THVR and LRTHR protocols have
similar performance in the average end-to-end delay. The performance of LRTHR is better when the algorithms are compared in the worst-case delays. Performance of SPEED is poor in both the average and worst case delays. In LRTHR, paths from source to sink will be shorter due to the dynamic velocity, two-hop information and some variation in the delays because of link reliability, THVR will select path based only on two-hop routing information.

\begin{figure}
\resizebox{7cm}{5cm}{\includegraphics{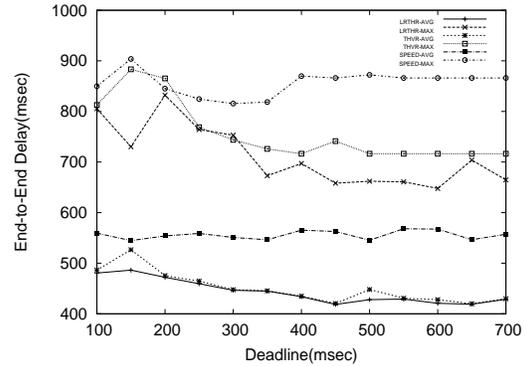}}
\begin{center}
\begin{small}
\caption{Average and Worst Case Delay Vs. Deadline}
\end{small}
\end{center}
\end{figure}
\vskip 2mm
Additionally, we examine the performance of the protocols under different loads. The number of sources are increased from 6 to 13, 
while the deadline requirement is fixed at 350 ms. Each source generates a CBR flow of 1 packet/second with a packet size of 150 bytes. 
From Figure 5 and Figure 6 it is observed that the DMR and ECPP plots ascend as the number of sources increase. The increase is
resulted by the elevated channel busy probability, packet contentions at MAC and network congestion by the increased 
number of sources and resulting traffic. Examination of the plots illustrates that LRTHR protocol has
lower DMR and also lower energy consumption per successfully transmitted packet.
\vskip 2mm
Figure 7 examines the packet end-to-end average and worst-case delays respectively. It is observed all the three protocols have
similar performance in the average and worst case end-to-end delay, till the number of sources are 10. The performance of LRTHR is better because the algorithm is able to spread the routes to the destination, since greater number of source nodes help in finding
links with more reliability in alternate paths and also provides better energy utilization.

%From Fig
%are compared in Fig
%are observed in Fig
%as shown in Fig

\begin{figure}
\resizebox{7cm}{5cm}{\includegraphics{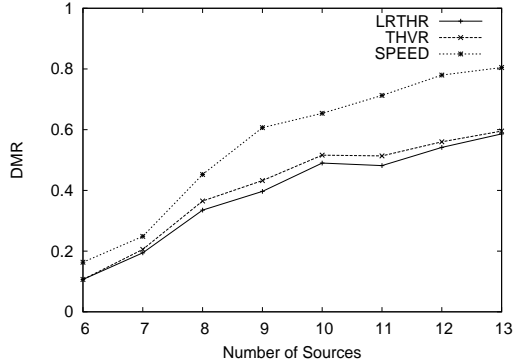}}
\begin{center}
\begin{small}
\caption{DMR Vs. No. of Sources}
\end{small}
\end{center}
\end{figure}

%Simulation results reveal that there is a reduction in DMR, ECP and end-to-end delays by the application of LRTHR algorithm. Overall link reliability, two-hop information and dynamic velocity enable the protocol to provide real time QoS support in WSNs.

\begin{figure}
\resizebox{7cm}{5cm}{\includegraphics{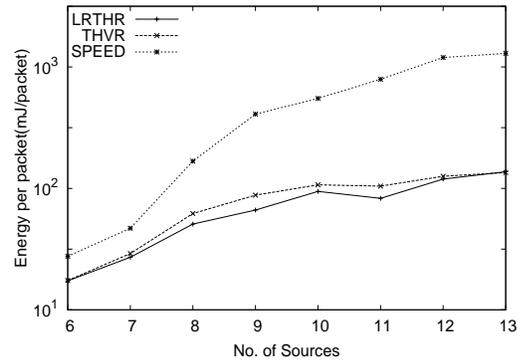}}
\begin{center}
\begin{small}
\caption{ECPP Vs. No. of Sources}
\end{small}
\end{center}
\end{figure}

Last, we study the performance of the residual energy cost function, the packet deadline is relaxed to a large value. Hence, when many
nodes can provide the required velocity, a node that has high residual energy can be chosen as a forwarding node. This will result in uniform load balancing among the nodes of the network.

\begin{figure}
\resizebox{7cm}{5cm}{\includegraphics{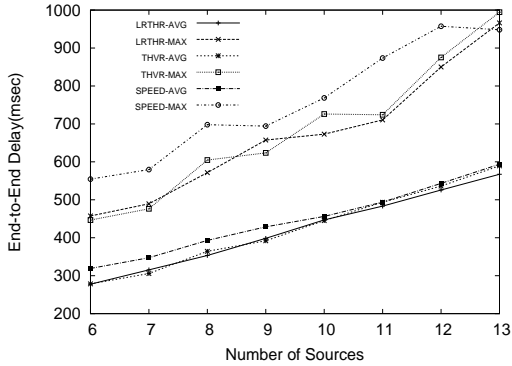}}
\begin{center}
\begin{small}
\caption{Average and Worst Case Delay Vs. No. of Sources}
\end{small}
\end{center}
\end{figure}

There are totally 200 nodes including 4 source nodes. The deadline is set to a large value of 600 ms. In THVR, the weighting factor $C$ 
is set at 0.7 to have a larger weighting on residual energy, in the proposed protocol we set weighting factors ($A$, $B$, $C$) to (0.1, 0.7, 0.2). Figure 8 and Figure 9 show the node energy consumption distribution in THVR and LRTHR respectively after 200 runs. 
As observed in THVR, some nodes along the path from
sources to sink are frequently chosen as forwarders and consume much more energy than the other, while in LRTHR only nodes close to the sources and sink consume relatively high energy. The latter is normal and inevitable especially as there may not be many optimal forwarding options near the sources and sink. Besides, by comparing Figure 8 to Figure 9, energy consumption in LRTHR is more evenly distributed among those between source and sink. The link reliability cost function will further aid to spread the routes to the destination compared to THVR. It can be observed that LRTHR will have a longer system lifetime due to the balancing.

\begin{figure}
\resizebox{7cm}{5cm}{\includegraphics{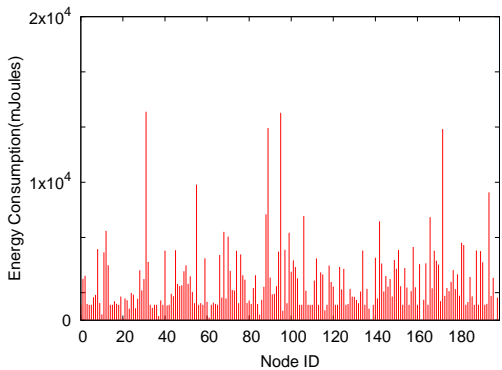}}
\begin{center}
\begin{small}
\caption{Node Energy Consumption in THVR}
\end{small}
\end{center}
\end{figure}

\begin{figure}
\resizebox{7cm}{5cm}{\includegraphics{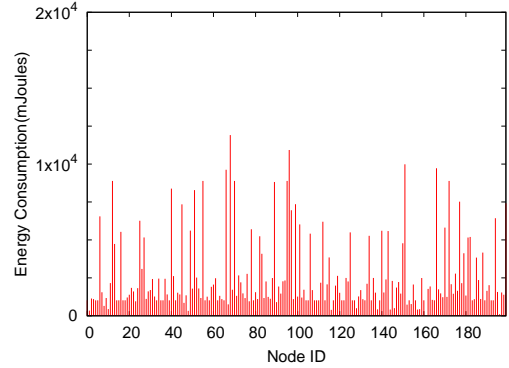}}
\begin{center}
\begin{small}
\caption{Node Energy Consumption in LRTHR}
\end{small}
\end{center}
\end{figure}

\section{CONCLUSIONS}
\label{section:conclusions}
In this paper, we propose a link reliability based two-hop neighborhood based quality of service (QoS) routing protocol for WSN.
Our proposed protocol is different from THVR, as it considers reliability and dynamic velocity that can be adjusted for each packet according to the required deadline. It balances the load only among nodes estimated to offer the required QoS.  The LRTHR protocol is able to augment real-time delivery by an able integration of link reliability, two-hop information and dynamic velocity. Future work 
can be carried out to support differentiated service and consider transmission power as a metric in forward node selection.

%The protocol is able to reduce the DMR, end-to-end delay and improvement
%on energy consumption balance throughout the network.

%

\small
\balance

\noindent{\includegraphics[width=1in,height=1.7in,clip,keepaspectratio]{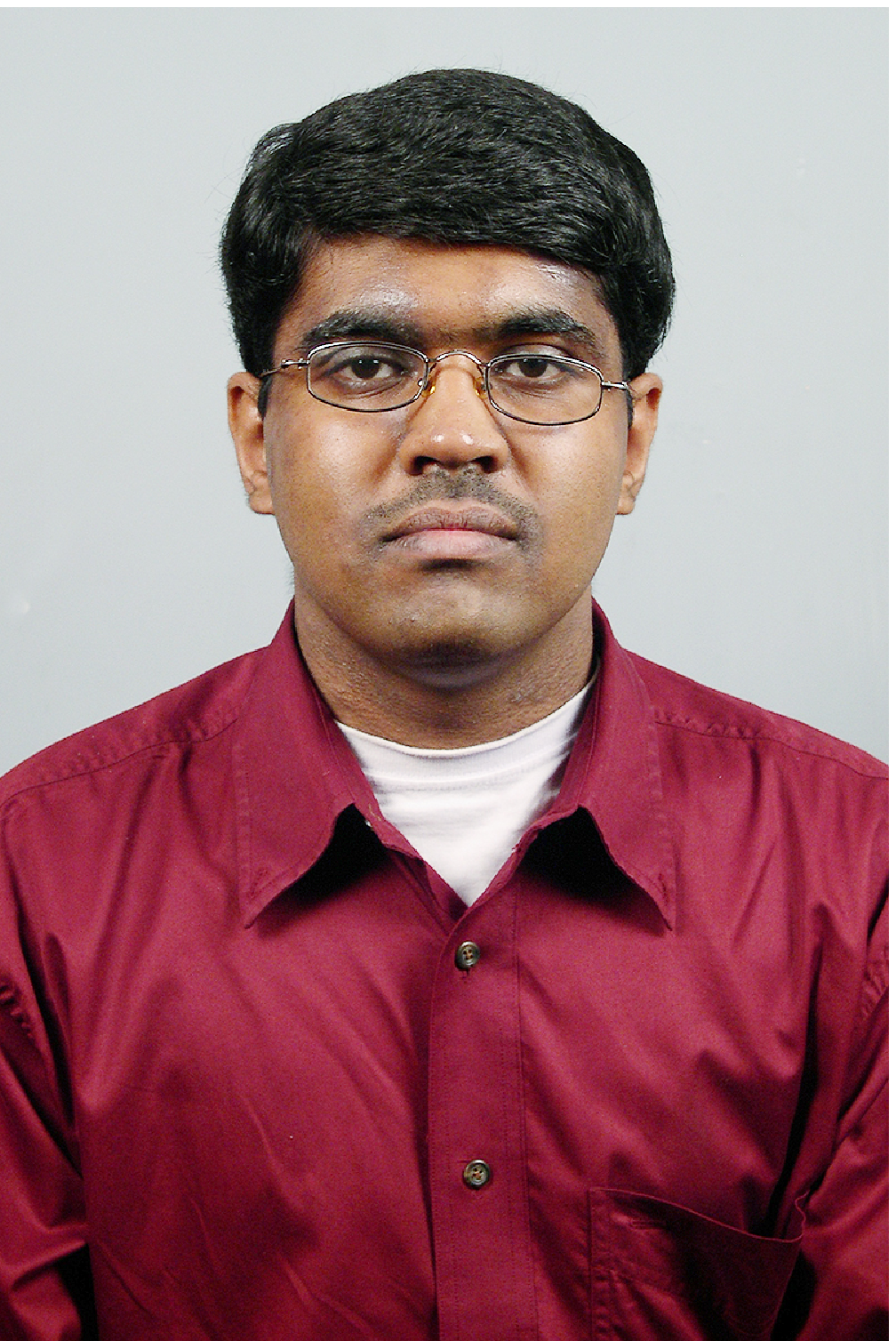}}
\begin{minipage}[b][1in][c]{1.8in}
{\centering{\bf {T Shiv Prakash}} is an Assistant Professor in the Department of Computer Science and Engineering at Vijaya Vittala Institute of Technology, Bangalore, India. He obtained his B.E and M.S Degrees in Computer Science and Engineering from Bangalore University, Bangalore. He is presently pursuing his Ph.D. pr-}\\\\\\
\end{minipage}\\
ogramme in the area of Wireless Sensor Networks in Bangalore University. His research interest is in the area of Sensor Networks, Embedded Systems and Digital Multimedia. \\\\

%\noindent{\includegraphics[width=1in,height=1.7in,clip,keepaspectratio]{kbr.eps}}
%\begin{minipage}[b][1in][c]{1.8in}
%{\centering{\bf {K B Raja}} is currently the xxxxx, University Visvesvaraya College of Engineering, Bangalore University, lore. He obtained his Bachelor of Engineering from University Visvesvaraya College of Engineering. He rece-}\\\\
%\end{minipage}
%xxxxxxxxxxxxxxxxxxxxxxxxxxxxxxxxxxxxxxxxxxxxxxxxxxxxxxxxxxxxxxxxxxxxxxxxxxxxxxxxxxxxxxxxxxxxxxxxxxxxxxx
%xxxxxxxxxxxxxxxxxxxxxxxxxxxxxxxxxxxxxxxxxxxxxxxxxxxxxxxxxxxxxxxxxxxxxxxxxxxxxxxxxxxxxxxxxxxxxxxxxxxxxxxxx\\\\
\noindent{\includegraphics[width=1in,height=1.7in,clip,keepaspectratio]{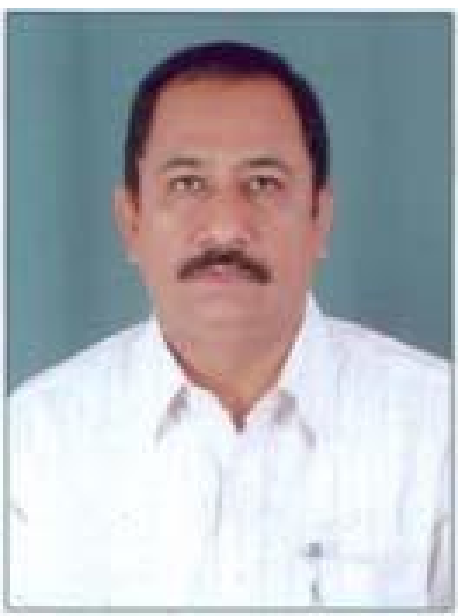}}
\begin{minipage}[b][1in][c]{1.8in}
{\centering{\bf {K B Raja }} is an Associate Professor, Dept. of Electronics and Communication Engg, University Visvesvaraya college of Engg, Bangalore University, Bangalore. He     obtained his Bachelor of Engineering and Master of Engineering in Electronics and Communication  En- }\\\\
\end{minipage}
gineering from University   Visvesvaraya  College     of Engineering, Bangalore. He was awarded
Ph.D. in Computer Science and Engineering from Bangalore
University. He has over 100 research publications in refereed
International Journals and Conference Proceedings. His research
interests include Image Processing, Biometrics, VLSI      Signal
Processing, Computer Networks.
 \\\\

\noindent{\includegraphics[width=1in,height=1.8in,clip,keepaspectratio]{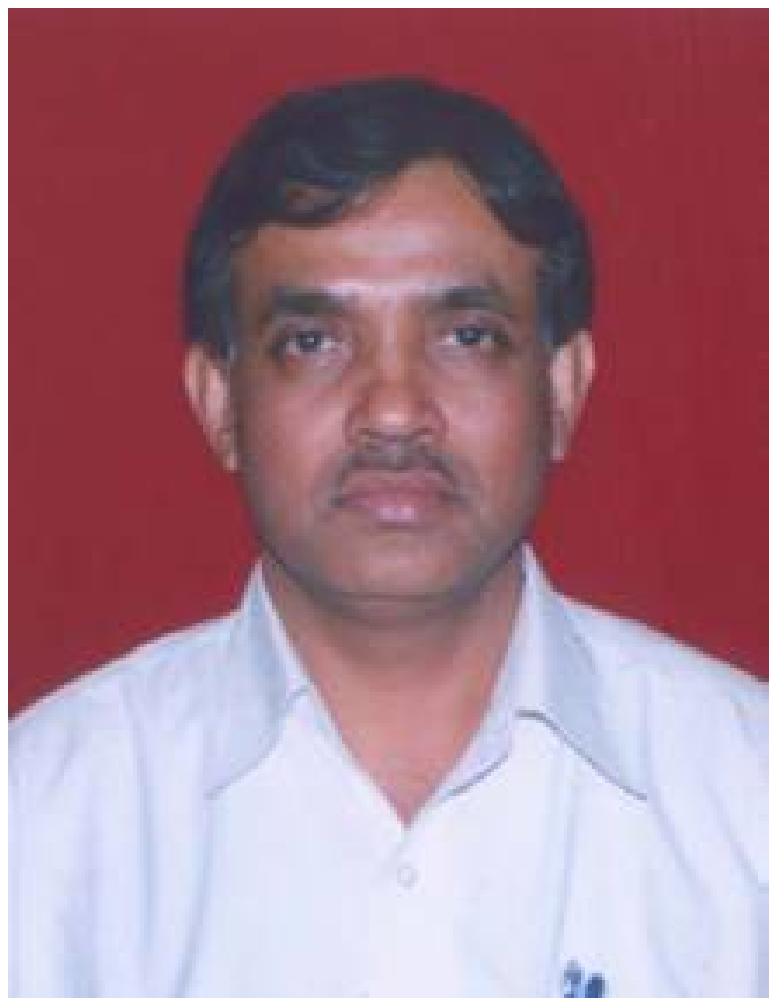}}
\begin{minipage}[b][1in][c]{1.8in}
{\centering{\bf {Venugopal K R}} is currently the Principal, University Visvesvaraya College of Engineering, Bangalore University, Bangalore. He obtained his Bachelor of Engineering from University Visvesvaraya College of Engineering. He received his Masters degree in Computer Science and} \\ \\
\end{minipage}
Automation from Indian Institute of Science Bangalore. He was awarded Ph.D. in Economics from Bangalore University and Ph.D. in Computer Science from Indian Institute of Technology, Madras. He has a distinguished academic career and has degrees in Electronics, Economics, Law, Business Finance, Public Relations, Communications, Industrial Relations, Computer Science and Journalism. He has authored and edited 35 books on Computer Science and Economics, which include Petrodollar and the World Economy, C Aptitude, Mastering C, Microprocessor Programming, Mastering C++ and Digital Circuits and Systems etc.. During his three decades of service at UVCE he has over 300 research papers to his credit. His research interests include Computer Networks, Wireless Sensor Networks, Parallel and Distributed Systems, Digital Signal Processing and Data Mining.\\ \\

\noindent{\includegraphics[width=1in,height=1.6in,clip,keepaspectratio]{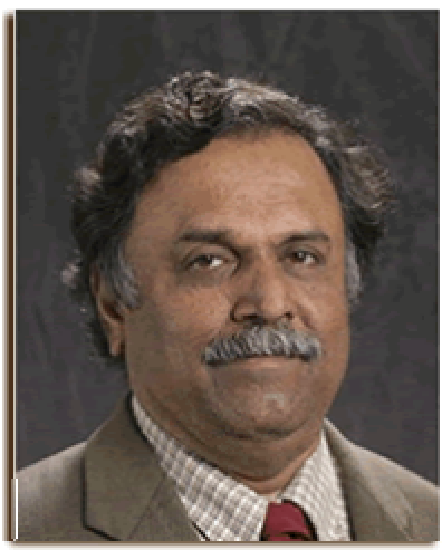}}
\begin{minipage}[b][1in][c]{1.8in}
{\centering{\bf {S S Iyengar}} is currently the Roy Paul Daniels Professor and Chairman of the Computer Science Department at Louisiana State University. He heads the Wireless Sensor Networks Laboratory and the Robotics Research Laboratory at LSU.}\\\\
\end{minipage}
He has been involved with research in High Performance Algorithms, Data Structures, Sensor Fusion and Intelligent Systems, since receiving his Ph.D degree in 1974 from MSU, USA. He is Fellow of IEEE and ACM. He has directed over 40 Ph.D students and 100 Post Graduate students, many of whom are faculty at Major Universities worldwide or Scientists or Engineers at National Labs/Industries around the world. He has published more than 500 research papers and has authored/co-authored 6 books and edited 7 books. His books are published by John Wiley \& Sons, CRC Press, Prentice Hall, Springer Verlag, IEEE Computer Society Press etc.. One of his books titled Introduction to Parallel Algorithms has been translated to Chinese.\\\\

\noindent{\includegraphics[width=1in,height=1.8in,clip,keepaspectratio]{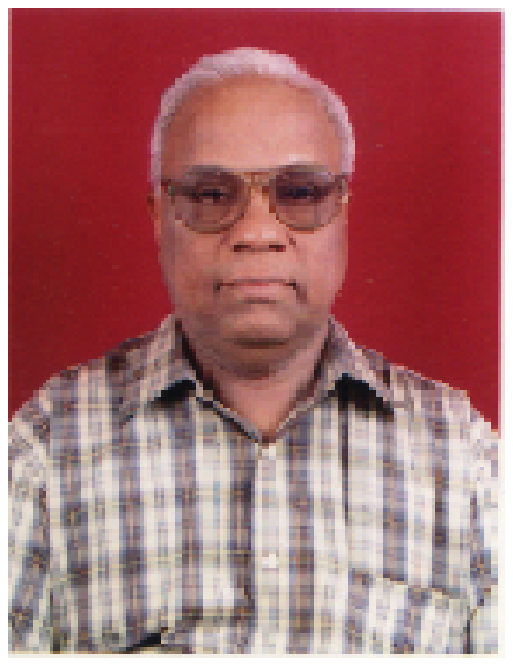}}
\begin{minipage}[b][1in][c]{1.8in}
{\centering{\bf {L M Patnaik}} is currently Honorary Professor, Indian Institute of Science, Bangalore, India. He was a Vice Chancellor, Defense Institute of Advanced Technology, Pune, India and was a Professor since 1986 with the Department of Computer Science and Automation, Indian} \\ \\
\end{minipage}
Institute of Science, Bangalore. During the past 35 years of his service at the Institute he has over 700 research publications in refereed International Journals and refereed International Conference Proceedings. He is a Fellow of all the four leading Science and Engineering Academies in India;  Fellow of the IEEE and the Academy of Science for the Developing World. He has received twenty national and international awards; notable among them is the IEEE Technical Achievement Award for his significant  contributions to High Performance Computing and Soft Computing. His areas of research interest have been Parallel and Distributed Computing, Mobile Computing, CAD for VLSI circuits, Soft Computing and Computational Neuroscience.
\balance
\end{document}